\documentclass{article}
\usepackage{amsmath,amssymb,amsfonts}
\usepackage{bbm}
\usepackage{graphicx}
\usepackage{hyperref}
\usepackage{braket}
\usepackage{mathtools}
\usepackage{amsthm}
\usepackage{authblk}

\newtheorem*{theorem*}{Theorem}
\newtheorem*{lemma*}{Lemma}

\usepackage{amsmath}
\usepackage[sorting=none]{biblatex}
\title{Unveiling Symmetries Patterns: A Study of Circular and Linear Harmonic Oscillator Chains}
\author[1]{Edoardo Spezzano\thanks{Email: e.spezzano@studenti.unipi.it}}
\author[1]{Alberto Iommi\thanks{Email: a.iommi1@studenti.unipi.it}}
\affil[1]{\textit{Dipartimento di Fisica E. Fermi, Università di Pisa, Largo B. Pontecorvo 3, I-56127 Pisa, Italy}}

\date{}

\begin{document}

\maketitle

\begin{abstract}
The purpose of this article is the study of the symmetries in a circular and linear harmonic oscillator chains system, and consequently use them as a means to find the eigenvalues of these configurations. Furthermore, a hidden $\mathbb{Z}_2$ group structure arises in both problems, showing how a degenerate spectrum in the circular case is attributable to the specific geometry producing a $\mathbb{Z}_N$ symmetry.

\end{abstract}

\section{Introduction}
The investigation of normal modes in complex physical systems has long fascinated physicists and mathematicians alike. These normal modes represent natural oscillatory configurations, shedding light on its dynamic behavior and underlying symmetries. This article delves into two fundamental problems in classical physics, each offering profound insights into the world of oscillations.\\

The objective of this work is to explore and analyze the frequency spectra associated with two classical systems: the harmonic oscillator chains in both circular and linear set-ups. In each scenario, identical masses are interconnected by springs, all possessing the same spring constant. Despite their apparent differences, these configurations share common ground in the context of oscillation theory.\\

By analysing the frequency spectra of these systems, we aim to deepen our understanding of their intrinsic behavior, unveil the underlying symmetries, and explore their significance in the broader landscape of classical physics.

\section{Problem Statement}
Before we delve into the analysis of frequency spectra, we have to establish the fundamental mathematical equations and boundary conditions valid for both problems. We begin by examining the system's Lagrangian \cite{goldstein2001classical, kibble2004classical}:

\begin{equation}
    \mathcal{L} = \sum_{i=1}^{N} \frac{1}{2} m \dot{x}_{i}^2 - \mathcal{U}_{\rm int}
\end{equation}

where $x_i$ represent the coordinates of the respective particles and $\mathcal{U}_{\rm int}$ represents the interaction potential. We can obtain the equations of motion using the Euler-Lagrange equations, resulting in:

\begin{equation}
    m \ddot{x}_{i} = -\frac{\partial \mathcal{U}_{\rm int}}{\partial x_i}
\end{equation}

Assuming the potential energy to be quadratic in terms of $x_i$, we can simplify the problem to the following form:

\begin{equation}
    \ddot{x}_i = \omega_0^2 H_{ij} x_j , \quad \omega_0^2 \equiv \frac{k}{m}
\end{equation}

with $k$ representing the spring constant and $H$ a real $N \times N$ matrix. The solutions to this equation are linear combinations of the normal modes, which correspond to oscillations at well-defined frequencies. In fact, we can express them in the form $x_i(t) = x_{i}^{0} e^{i \omega t}$, which, when substituted into the previous equation, yields an eigenvalue equation:

\begin{equation}
    \lambda x_{i}^{0} = H_{ij} x_{j}^{0} , \quad \lambda \equiv -\frac{\omega^2}{\omega_0^2}
\end{equation}

In the following sections of this article we will delve into the methods for finding the eigenvalues $\lambda$.

\section{Theoretical Methods and Analytical Framework}
As we have seen, the key to our analysis lies on finding the eigenvalues of the matrix $H$. We will try to find them on both cases, using a group theory approach.

\subsection*{Normal Modes in a Circular Chain}
The matrix $H_c$ for a circular chain in $N$ dimension takes the following form:

\begin{equation}
\begin{aligned}
H_c = \begin{pmatrix}
-2 & 1 & 0 & \cdots & 0 & 0 & 1 \\
1 & -2 & 1 & 0 & \cdots & 0 & 0 \\
0 & 1 & -2 & 1 & 0 & \cdots & 0 \\
\vdots & \ddots & \ddots & \ddots & \ddots & \ddots & \vdots \\
0 & \cdots & 0 & 1 & -2 & 1 & 0 \\
0 & 0 & \cdots & 0 & 1 & -2 & 1 \\
1 & 0 & 0 & \cdots & 0 & 1 & -2 \\
\end{pmatrix}_{N \times N}
\end{aligned}
\end{equation}

We can express the matrix $H_c$ as:

\begin{equation}
\begin{aligned}
H_c = T + T^{-1} - 2 \mathbbm{1}
\end{aligned}
\end{equation}

where $T$ is defined as:

\begin{equation}
\begin{aligned}
T = \begin{pmatrix}
0 & 0 & 0 & \cdots & 0 & 0 & 1 \\
1 & 0 & 0 & \cdots & 0 & 0 & 0 \\
0 & 1 & 0 & \cdots & 0 & 0 & 0 \\
\vdots & \ddots & \ddots & \ddots & \vdots & \vdots & \vdots \\
0 & \cdots & 0 & 1 & 0 & 0 & 0 \\
0 & 0 & \cdots & 0 & 1 & 0 & 0 \\
0 & 0 & 0 & \cdots & 0 & 1 & 0 \\
\end{pmatrix}_{N \times N}
\end{aligned}
\end{equation}

We can observe that $T$ is a matrix who shifts mass positions by 1. Indeed, this operator corresponds to a symmetry of the problem, which can be demonstrated through the equation:

\begin{equation}
[H_c,T]=0
\end{equation}

Furthermore, it is evident that this matrix constitutes a representation of the group  $\mathbb{Z}_N$ \cite{georgi1982lie} (also commonly referred to as the \textit{regular} representation in the literature). Consequently, its eigenvalues are determined by $\lambda^N=1$, resulting in $\lambda_k=e^{i\frac{2\pi k}{N}}$ for $k=0, \dots , N-1$. Thus, the spectrum is given by:

\begin{equation}
\begin{aligned}
\lambda = \lambda_k + \frac{1}{\lambda_k} - 2
\end{aligned}
\end{equation}

and substituting the $\lambda_k$ expression, we obtain:

\begin{equation}\label{circular}
\begin{aligned}
\lambda= 2 \cos \frac{2k\pi}{N} - 2 = -4 \sin^2 \frac{k\pi}{N} 
\end{aligned}
\end{equation}

who leads to the well-known solution:

\begin{equation}
\begin{aligned}
\omega(k) = 2 \omega_0 \sin \frac{k\pi}{N}, \quad k \in [-N + 1,  N - 1]
\end{aligned}
\end{equation}

We can observe that the formula implies a degeneracy on the spectrum caused by the presence of an additional symmetry in the problem. Specifically, one can see that the so-called exchange matrix \cite{Horn:2012:MA}:

\begin{equation}
\begin{aligned}
J = \begin{pmatrix}
0 & 0 & \cdots & 0 & 1 \\
0 & \cdots & 0 & 1 & 0 \\
\vdots & \reflectbox{$\ddots$} & \reflectbox{$\ddots$} & 0 & \vdots \\
0 & 1 & \reflectbox{$\ddots$} & \vdots & 0 \\
1 & 0 & \cdots & 0 & 0 \\
\end{pmatrix}_{N \times N}
\end{aligned}
\end{equation}

satisfies $[H_c,J] = 0$, but $[J,T] \neq 0$ \footnote{More precisely, this occurs for $N \geq 3$. Indeed for the case of $N=2$, we have $J=T$.}, showing that, as just said, the spectrum is necessarily degenerate (see Theorem \cite{1326}). Additionally, another intriguing property of this operator is that $J^2= \mathbbm{1}$ and so constitute a representation of the group $\mathbb{Z}_2$. Taking all this result into account, we can see that the symmetry group in the case of a circular chain is $ \mathbb{Z}_N \times \mathbb{Z}_2$.

\subsection*{Normal Modes in a Linear Chain}
In the case of a linear chain, the matrix $H_l$ takes the following form:

\begin{equation}
\begin{aligned}
H_l = \begin{pmatrix}
-2 & 1 & 0 & \cdots & 0 & 0 & 0 \\
1 & -2 & 1 & 0 & \cdots & 0 & 0 \\
0 & 1 & -2 & 1 & 0 & \cdots & 0 \\
\vdots & \ddots & \ddots & \ddots & \ddots & \ddots & \vdots \\
0 & \cdots & 0 & 1 & -2 & 1 & 0 \\
0 & 0 & \cdots & 0 & 1 & -2 & 1 \\
0 & 0 & 0 & \cdots & 0 & 1 & -2 \\
\end{pmatrix}_{N \times N}
\end{aligned}
\end{equation}

Similarly to the circular example, we can inquire whether the system exhibits symmetries or not. It is not difficult to see that in this case too, we have $[H_l,J] = 0$, indicating that the system possesses at least a $\mathbb{Z}_2$ symmetry. To understand the full symmetry group present, initially notice that the following result holds:

\begin{lemma*}
Given a $N \times N $ matrix $\mathcal{O}$, it holds:
\begin{equation}\label{teo}
    [H_l, \mathcal{O}] = 0 \Leftrightarrow \mathcal{O} = \sum_{n=0}^{N-1} c_n U_n \left( \frac{\mathcal{H}}{2} \right), \quad \mathcal{H} = H_l + 2 \mathbbm{1},
\end{equation}
where the functions $U_n(x)$ are the Chebyshev polynomials of the second kind \cite{mason2002chebyshev}.
\end{lemma*}

For a proof, refer to the appendix (\ref{app}).\\
As a corollary of this lemma, we see that the symmetry group is certainly abelian. Therefore, by the classification theorem for finite abelian groups \cite{isaacs1994finite}, our attention is drawn to $\mathbbm{Z}_n$. However, it becomes evident that we cannot have a symmetry group with $n \geq 3$ because $[H_l,T] \neq 0$. Thus, the only symmetry group is generated by $J$, which is $\mathbbm{Z}_2$.\\

In conclusion, let us explore a method to determine the spectrum. This involves deriving the following recurrence relation found in the characteristic polynomials:

\begin{equation}\label{charac}
\begin{aligned}
P_N(\lambda) = \lambda P_{N-1}(\lambda) - P_{N-2}(\lambda)
\end{aligned}
\end{equation}

This recurrence relation is similar to the one satisfied by the Chebyshev polynomials cited above. Specifically, the connection between them is as follows:

\begin{equation}
\begin{aligned}
P_{N}(x) = U_N\left(\frac{x}{2}\right)
\end{aligned}
\end{equation}

From the above equation, we notice that solving the secular equation $P_N(x) = 0$ coincides with the roots of Chebyshev polynomials, which are given by:

\begin{equation}
\begin{aligned}
x_k = \cos\left(\frac{k \pi}{N+1} \right), \quad k = 0, \ldots, N - 1
\end{aligned}
\end{equation}

Hence, the eigenvalues of our problem are determined as:

\begin{equation}\label{linear}
\begin{aligned}
\lambda = 2 \left[\cos\left(\frac{k \pi}{N+1}\right) - 1\right] = -4 \sin^2\left(\frac{k \pi}{2(N+1)}\right)
\end{aligned}
\end{equation}

leading to:

\begin{equation}
\begin{aligned}
\omega(k) = 2 \omega_0 \sin\left(\frac{k \pi}{2(N+1)}\right), \quad k \in [-N + 1,  N - 1]
\end{aligned}
\end{equation}

It is worth noting that the spectrum is very similar to that obtained in the case of the circular chain. This similarity is discussed in more detail in the next section.

\subsection*{Note on the Anti-Commutator}
Notably, the diagonal matrix:

\begin{equation}
\begin{aligned}
S = \begin{pmatrix}
1 & 0 & 0 & 0 & \cdots & 0 \\
0 & -1 & 0 & 0 & \cdots & 0 \\
0 & 0 & 1 & 0 & \cdots & 0 \\
0 & 0 & 0 & -1 & \ddots & \vdots \\
\vdots & \vdots & \vdots & \ddots & \ddots & 0 \\
0 & 0 & 0 & \cdots & 0 & (-1)^{N+1} \
\end{pmatrix}_{N \times N}
\end{aligned}
\end{equation}

in the case of a linear chain satisfies the following relationship:

\begin{equation}
\begin{aligned}
\{ H_l, S \} = -4 S .
\end{aligned}
\end{equation}

So, due to the anti-commutation property, if $\lambda$ is an eigenvalue of $H$, then $-4-\lambda$ is necessarily also an eigenvalue, as expressed by:

\begin{equation}
\begin{aligned}
H_l(S \ket{n}) = (-\lambda-4)( S \ket{n})
\end{aligned}
\end{equation}

This relationship is also valid in the case of the circular chain with $N$ even. In fact, it can be easily shown that $\{H_{c,2N}, S_{2N}\}=-4 S_{2N}$. This result is anticipated by examining the two spectra given by the equations \eqref{circular} and \eqref{linear}.

\section{Conclusions}

As for the circular chain, we identified a $\mathbb{Z}_n \times \mathbb{Z}_2$ symmetry, simplifying the eigenvalue problem resolution. This symmetry facilitated a more streamlined approach, employing tailored mathematical tools. In the case of the linear chain, we uncovered a $\mathbb{Z}_2$ symmetry and determined its spectrum, highlighting the crucial role of Chebyshev polynomials.\\

An intriguing feature emerges from the eigenvalues due to the inherent anti-symmetry relationship between the Hamiltonian operator ($H$) and the operator $S$, introducing distinctive patterns into the eigenvalue spectra. We observed an analogy between the linear chain and the even case of the circular chain, arising from the anti-commutation.
\appendix

\section*{Appendix: Proof of the Lemma}\label{app}
To demonstrate \ref{teo}, we first observe that proving the commutation of a given operator $M$ is equivalent to proving it for $\mathcal{H} = H_l + 2\mathbbm{1}$. This equivalence becomes evident from the following relation:

\begin{equation}
    [\mathcal{H}, M] = 0 \iff [H_l, M] = 0
\end{equation}

Now, considering the form of $\mathcal{H} = \delta_{i-1,j} + \delta_{i+1,j}$ and the equation above, it follows that the matrix $M$ must satisfy:

\begin{equation}
    M_{i-1,j} + M_{i+1,j} = M_{i,j-1} + M_{i,j+1}
    \label{eq1}
\end{equation}

This property is noteworthy, in particular it implies that:

\begin{itemize}
    \item $M_{ij} = M_{ji}$, \quad \text{symmetry with respect to the main diagonal}
    \item $M_{ij} = M_{N+1-j,N+1-i}$, \quad \text{symmetry with respect to the antidiagonal}
\end{itemize}

Both of these conditions can be proven by examining the previous equations at the respective matrix vertices.\\

Given these conditions, it is evident that the dimension of the vector space is significantly smaller than the space of symmetric matrices; in particular, it is easy to see that it is equal to $N$.\\

At this point, the idea is to construct a basis for this vector space. It becomes evident that the previous equations impose strong constraints on the construction of matrices, leaving only one possible approach. For instance, when attempting to construct the first element of the basis as follows:

\begin{equation}
\begin{pmatrix}
1 & 0 & 0 & \cdots & 0 \\
* & * & * & * & * \\
* & * & * & * & * \\
* & * & * & * & * \\
* & * & * & * & * \\
\end{pmatrix}
\end{equation}

it is necessary to construct the identity matrix. The same holds true for the second element; in fact, aiming to construct:

\begin{equation}
\begin{pmatrix}
0 & 1 & 0 & \cdots & 0 \\
* & * & * & * & * \\
* & * & * & * & * \\
* & * & * & * & * \\
* & * & * & * & * \\
\end{pmatrix}
\end{equation}

one is constrained to create the matrix $\mathcal{H}$. In general, it is not difficult to realize that, given $N$, a potential basis is provided by the following set: $\{ P_0(\mathcal{H}), P_1(\mathcal{H}), \ldots, P_{N-1}(\mathcal{H})\}$, where

\begin{equation}
P_{N}(x) = \sum_{k=0}^{\lfloor N/2 \rfloor} (-1)^k \binom{N-k}{k} x^{N-2k}
\end{equation}

It turns out that this is actually a way to represent the second type of Chebyshev polynomials, and the connection is expressed as $P_n(2x)=U_n(x)$. These polynomials are known for being a set of polynomials that are all independent of each other.\\

In summary, if one wishes to construct a matrix $M$ that commutes with $\mathcal{H}$, it must necessarily be written as:

\begin{equation}
    M = \sum_{n=0}^{N-1} a_n P_n(\mathcal{H})
\end{equation}

This completes the proof in the $\Rightarrow$ direction. The proof in the $\Leftarrow$ direction is straightforward.\\

It is worth noting that $P_{N}(\mathcal{H})=0$ as a consequence of the Cayley-Hamilton theorem \cite{cayley1858theorem} (see \eqref{charac}), providing a consistency check to the lemma.

\section*{Acknowledgments}
We would like to extend our heartfelt gratitude to Daniel Loni for meticulously reading the article and offering invaluable feedback and insights. His careful review significantly enhanced the quality of our work.

\printbibliography

@book{goldstein2001classical,
  title={Classical mechanics},
  author={Goldstein, Herbert and Poole, Charles and Safko, John},
  year={2001},
  publisher={Addison-Wesley}
}

@book{kibble2004classical,
  title={Classical mechanics},
  author={Kibble, Tom W and Berkshire, Frank H},
  year={2004},
  publisher={Imperial College Press}
}

@book{georgi1982lie,
  title={Lie algebras in particle physics},
  author={Georgi, Howard},
  year={1982},
  publisher={Westview Press}
}

@book{mason2002chebyshev,
  title={Chebyshev polynomials},
  author={Mason, John C and Handscomb, David C},
  year={2002},
  publisher={CRC Press}
}

@book{Horn:2012:MA,
 author = {Horn, Roger A. and Johnson, Charles R.},
 title = {Matrix Analysis},
 edition = {2},
 publisher = {Cambridge University Press},
 year = {2012},
 isbn = {9781139788885}
}

@article{cayley1858theorem,
  title={On the theory of linear transformations},
  author={Cayley, Arthur},
  journal={Cambridge Mathematical Journal},
  volume={X},
  pages={267--271},
  year={1858}
}

@book{1326,
  author = {Shankar, R},
  title = {Principles of quantum mechanics},
  publisher = {Springer},
  year = {1994},
  address = {New York},
  edition = {2nd ed.}
}

@book{isaacs1994finite,
  title={Finite Group Theory},
  author={Isaacs, I. Martin},
  year={1994},
  publisher={American Mathematical Society}
}

\end{document}